\documentclass[]{emulateapj}
\usepackage{apjfonts}
\newcommand{\be}{\begin{equation}}
\newcommand{\ee}{\end{equation}}
\newcommand{\bea}{\begin{eqnarray}}
\newcommand{\eea}{\end{eqnarray}}
\newcommand{\Msun}{M_{\odot}}

\def\Mdm{\ M_{\rm DM}}
\def\Msun{\ M_\Sol}

\shortauthors{CONROY, LOEB \& SPERGEL}
\shorttitle{EVIDENCE AGAINST DARK HALOS SURROUNDING GLOBULAR CLUSTERS}

\begin{document}
\journalinfo{The Astrophysical Journal}

\title{Evidence Against Dark Matter Halos Surrounding the Globular
  Clusters MGC1 and NGC 2419}

\author{Charlie Conroy\altaffilmark{1}, Abraham Loeb\altaffilmark{1},
  \& David N. Spergel\altaffilmark{2}}

\altaffiltext{1}{Harvard-Smithsonian Center for Astrophysics,
  Cambridge, MA, USA} 
\altaffiltext{2}{Department of Astrophysical
  Sciences, Princeton University, Princeton, NJ, USA}

\begin{abstract}

  The conjecture that the ancient globular clusters (GCs) formed at
  the center of their own dark matter halos was first proposed by
  \citet{Peebles84}, and has recently been revived to explain the
  puzzling abundance patterns observed within many GCs.  In this paper
  we demonstrate that the outer stellar density profile of isolated
  GCs is very sensitive to the presence of an extended dark halo.  The
  GCs NGC 2419, located at 90 kpc from the center of our Galaxy, and
  MGC1, located at $\sim200$ kpc from the center of M31, are ideal
  laboratories for testing the scenario that GCs formed at the centers
  of massive dark halos. Comparing analytic models to observations of
  these GCs, we conclude that these GCs cannot be embedded within dark
  halos with a virial mass greater than $10^6\Msun$, or, equivalently,
  the dark matter halo mass-to-stellar mass ratio must be
  $\Mdm/M_\ast<1$.  If these GCs have indeed orbited within weak tidal
  fields throughout their lifetimes, then these limits imply that
  these GCs did not form within their own dark halos.  Recent
  observations of an extended stellar halo in the GC NGC 1851 are also
  interpreted in the context of our analytic models.  Implications of
  these results for the formation of GCs are briefly discussed.

\end{abstract}



\section{Introduction}
\label{s:intro}

Despite decades of intense theoretical effort, the formation of the
ancient globular clusters (GCs) remains a largely unsolved problem.
\citet{Peebles84} considered the possibility that GCs form within
their own dark matter (DM) halos at high redshift.  The growing
evidence for significant self-enrichment in GCs and the broad
acceptance of hierarchical structure formation has deepened interest
in this formation scenario.  Evidence against this scenario was found
in the observations of thin tidal tails surrounding many GCs
\citep[e.g.,][]{Grillmair95, Odenkirchen03}, because numerical
simulations showed that such tidal tails do not form if GCs reside
within extended halos \citep{Moore96b}.  However, later work
highlighted the fact that even if Milky Way (MW) GCs were once
embedded within massive dark halos, these halos would have been
tidally stripped away by the present epoch \citep{Bromm02,
  Mashchenko05b}.  This requires relatively strong tidal fields, which
suggests that GCs in the outer halo of the MW may still be embedded
within dark halos, if they formed within them.

Other theories for the formation of GCs do not appeal to formation at
the center of dark halos.  \citet{Fall85} proposed that GCs form from
thermal instabilities in the hot gaseous halos expected to surround
massive galaxies today.  This proposal suffers from the fact that many
galaxies that host GCs are not expected to reside in halos massive
enough to support a hot halo, such as dwarf spheroidals.

\citet{Gunn80} was the first to suggest that GCs could form in the gas
compressed by strong shocks.  This proposal received tentative
confirmation with the discovery of many massive young star clusters
within the interacting Antennae system \citep{Whitmore95, Whitmore99}
and the discovery of super star clusters within nearby galaxies
\citep[e.g.,][]{Holtzman92}.  This scenario, modified to include as
formation sites any massive, dense, cold patch of gas, is now the
prevailing paradigm for GC formation \citep[e.g.,][]{Harris94}, and,
when incorporated into our broader theory of cosmological structure
formation, is capable of explaining a variety of observations
\citep[e.g.,][]{Ashman92, Kravtsov05b, Muratov10}.

This prevailing paradigm for GC formation is complicated by the
existence of nuclear star clusters \citep{Boker04, Walcher05,
  Walcher06}, which implies that at least some GC-like systems can
form at the centers of massive dark halos.  The existence of {\it
  young} nuclear star clusters makes this point particularly
compelling, since these clusters could not have migrated to the center
via dynamical friction.  Thus, while dark halos are not {\it
  necessarily required} for GC formation, the conditions for GC
formation may sometimes be realized at the centers of dark halos.
Clearly, further constraints on the formation sites of GCs is
desirable.

In a series of papers, Spitzer and collaborators derived the kinematic
properties of stars in the stellar halo of a GC, where stars are only
marginally bound \citep{Spitzer71, Spitzer72}.  An important result
from this work was that the density profile of stars in the stellar
halo should scale as $r^{-3.5}$.  In the present work we build upon
these results by investigating the sensitivity of the stellar density
profile to the presence of a massive dark halo.

\section{The Stellar Halos of Globular Clusters}
\label{s:model}

\subsection{Analytic model}

In this section we derive the outer stellar density profile of GCs
embedded in a massive dark halo.  The following derivation closely
follows the assumptions and approximations made in a series of papers
by Spitzer and collaborators \citep{Spitzer71, Spitzer72, Spitzer87},
to which the reader is referred for details.

The density profile of a stellar system can be derived from its
distribution function, $f$, via:
\noindent
\be
n(r) \propto \int_{E<0}\, f(E,J)\,2\pi\,{\rm v}_t\,
{\rm d}{\rm v}_t\,{\rm d}{\rm v}_r,
\ee
\noindent
where ${\rm v}_t$ and ${\rm v}_r$ are the tangential and radial
velocities.  We assume that GC halo stars are on radial orbits, and
thus are justified in making the approximation that ${\rm v}^2={\rm
  v}_r^2$, and we can substitute ${\rm v}_t=J/r$.  Most importantly,
we assume that $f(E,J)=|E|\,g(J)$, where $g$ is some function of
angular momentum.  This functional form arises when the orbital
energies are only slightly below zero, the number of stars in the
system is large, and the system has reached a steady state
\citep[see][for details]{Spitzer72}.  These constraints require that
the two-body relaxation time is short compared to the age of the
Universe.  We then have:
\noindent
\be
n(r) \propto r^{-2}\,  g'(J)\, \int_{E<0}|E|\,{\rm dv},
\ee
\noindent
where $g'$ is some new function of angular momentum.  Assuming that
$J$ is not a function of $r$ in the stellar halo, we drop all
reference to $J$ from here on.

For a purely stellar system we have $E=\frac{1}{2}{\rm
  v}^2+\Phi_\ast$, where $\Phi_\ast$ is the potential of the stars and
is approximated by a Keplerian potential
($\Phi_\ast\propto-GM_\ast/r$).  Upon substitution into Equation 2 we
recover the familiar result that $n(r)\propto r^{-3.5}$ in the halo of
GCs.  This result has been confirmed by direct $N-$body simulations
\citep[e.g.,][]{Baumgardt02}.

Our task here is simply to re-evaluate this integral with the addition
of a DM potential, $\Phi_{\rm DM}$.  The distribution function of
weakly-bound stars is unchanged with the addition of a dark halo since
the derivation makes no reference to the form of the potential.  We
therefore have:
\noindent
\be
n(r) \propto r^{-2}\, \int_{E<0}\bigg|\frac{1}{2}{\rm v}^2+
\Phi_*+\Phi_{\rm DM} \bigg|{\rm dv},
\ee
\noindent
which upon integration becomes:
\noindent
\be
n(r) \propto  r^{-2}\, (  \Phi_\ast + \Phi_{\rm DM} )^{3/2}.
\ee
\noindent

\begin{figure}[!t]
\center
\resizebox{2.7in}{!}{\includegraphics{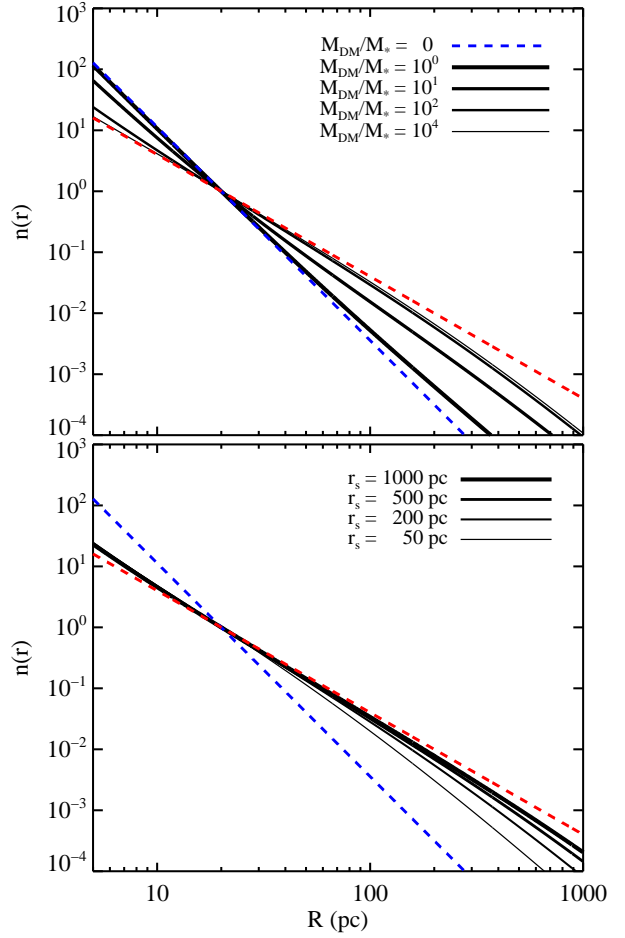}}
\vspace{0.1cm}
\caption{Stellar density profiles normalized to the density at 20 pc.
  Models are shown for several values of the dark halo-to-stellar mass
  ratio, $\Mdm/M_\ast$ ({\it top panel}) and dark halo scale radius,
  $r_s$ ({\it bottom panel}).  In the top panel $r_s=250$ pc, and in
  the bottom panel $\Mdm/M_\ast=10^2$.  The blue and red dashed lines
  have logarithmic slopes of $-3.5$ and $-2.0$, respectively.}
\label{fig:f1}
\vspace{0.1cm}
\end{figure} 

We assume an NFW density profile for the dark halo that is motivated
by collisionless $\Lambda$CDM cosmological simulations \citet{NFW96,
  NFW97}.  The implied dark halo potential is
\noindent
\be
\Phi_{\rm DM} = -G\Mdm g(c) \, \frac{{\rm ln}(1+r/r_s)}{r},
\ee
\noindent
where $\Mdm$ is the total dark halo `virial' mass, $c$ is the
concentration defined as $c\equiv r_{\rm v}/r_s$ where $r_{\rm v}$ is
the virial radius and $r_s$ is the scale radius, and $g(c)=[{\rm
  ln}(1+c)-c/(1+c)]^{-1}$.  Over the physically relevant range of
$2\lesssim c\lesssim10$, $g(c)$ varies from 2.3 to 0.7.

Finally then, we have the following expression for the stellar density
profile in the presence of a dark halo\footnote{The contribution from
  unbound stars is not included here.  We expect their contribution to
  be negligible because simulations consistently find that stars are
  unbound at a rate of $\sim1$\% per relaxation time.  Moreover, the
  density profile of the escapers is approximately $r^{-2}$
  \citep{Spitzer87}, even in the presence of a dark halo, and so their
  presence would not impact our conclusions.}:
\noindent
\be
\label{eqn:nr}
n(r) \propto r^{-3.5}\, \bigg[  1 +\frac{\Mdm}{M_\ast}
\,g(c)\, {\rm ln}(1+r/r_s) \bigg]^{3/2}.
\ee
\noindent

For $\Mdm/M_\ast\ll1$ we recover the familiar result of $n(r)\propto
r^{-3.5}$.  When the dark halo mass is significant, the profile can be
decomposed into three regimes.  At sufficiently small scales the first
term in brackets in equation \ref{eqn:nr} dominates over the second,
and the profile scales as $r^{-3.5}$.  At larger scales, the second
term dominates, and it takes on two limits for $r$ smaller or larger
than $r_s$.  For $r<r_s$ the second term scales as $r$ and the total
density profile then scales as $n(r)\propto r^{-2}$.  At scales
greater than $r_s$ the second term in brackets shallows, and the
resulting density profile consequently steepens.

\begin{figure}[!t]
\center
\resizebox{3.4in}{!}{\includegraphics{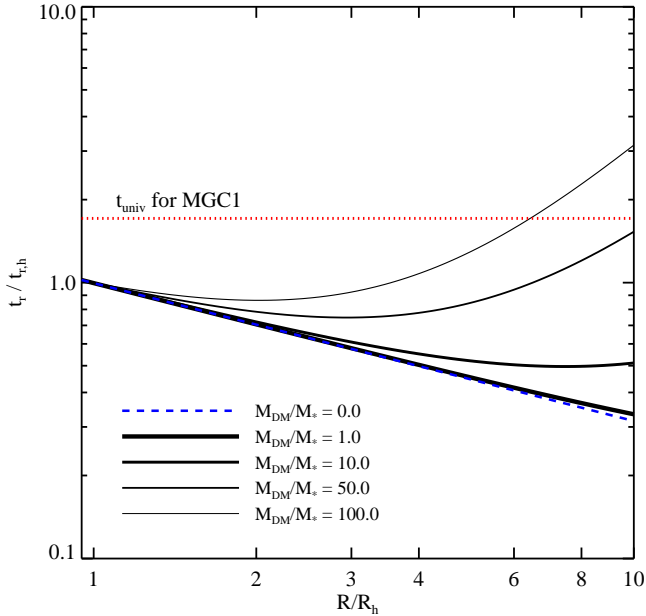}}
\caption{Relaxation time in units of the half-mass relaxation time, as
  a function of clustocentric distance.  Results are shown for several
  values of the dark matter-to-stellar mass ratio.  The age of the
  Universe is shown in units of the half-mass relaxation time of MGC1
  ({\it dotted line}).}
\label{fig:trelax}
\vspace{0.1cm}
\end{figure}

In Figure \ref{fig:f1} we show the expected stellar density profiles
for several values of the parameters $\Mdm/M_\ast$ and $r_s$.  For
simplicity, we have fixed the virial radius to $r_{\rm v}=1$ kpc
although the models are insensitive to this simplification.  Notice
the strong sensitivity to $\Mdm/M_\ast$ and the weak sensitivity to
the $r_s$ over the scales of interest. The weak sensitivity to $r_s$
is due to the fact that the logarithmic slope of the dark halo
potential varies slowly across $r_s$.

Figure \ref{fig:f1} demonstrates that the density profile over the
range $10\lesssim r\lesssim100$ pc is very sensitive to the presence
of a dark halo.  Our derivation of the density profile is strictly
appropriate only for the stellar halo of a GC, and so the profiles in
Figure \ref{fig:f1} will not represent real GCs on smaller scales.  We
have also ignored tidal stripping and the fact that the relaxation
time at large scales may under certain circumstances be longer than
the age of the Universe.

The derivation of the stellar density profile in the halo of GCs
relies on the assumption that the two-body relaxation time is short
compared to the age of the Universe.  We now verify under what
conditions this assumption is valid.  The radial dependence of the
effective relaxation time in the stellar halo can be estimated as
follows \citep[see][for details]{Lightman78}.  The relaxation time,
$t_r$, scales as
\noindent
\be
t_r \propto \frac{E^2}{D(\Delta E^2)},
\ee
where $D$ is the diffusion rate and $E$ is the energy.  The diffusion
rate is simply the energy change per unit time, which, for stars in
the halo, is \citep{Binney87}:
\be
D(\Delta E^2) \propto \frac{\epsilon^2}{P},
\ee
\noindent
where $P$ is the orbital period and $\epsilon$ is the (small) change
in energy per orbit.  The key feature of stars in the halo is that
they are on radial orbits that pass through the central regions of the
GC.  This fact implies that $\epsilon$ is approximately contant for
stars in the halo; i.e., the change in energy per orbit does not
depend on the apocentric distance of the orbit.  Thus:
\noindent
\be
t_r \propto E^2\,P.
\ee
\noindent
Assuming that the potential is a combination of a Keplerian and an NFW
dark halo, as we have throughout this section, we arrive at the
following expression for the relaxation time:
\noindent
\be
\label{e:tr}
t_r \propto \frac{1}{\sqrt{r}}\bigg[1+\frac{M_{\rm
    DM}}{M_\ast}\,g(c)\,\bigg({\rm ln}(1+r/r_s)-\frac{r/r_s}{1+r/r_s}\bigg)\bigg]^{3/2}.
\ee
\noindent
The relaxation time is a function of radius and dark matter-to-stellar
mass ratio.  These dependencies are illustrated in Figure
\ref{fig:trelax}.  In this figure the relaxation time is scaled to the
relaxation time at the half-mass radius.  Notice first the
counterintuitive result that in systems dominated by a Keplerian
potential the relaxation time in the halo is actually a {\it
  decreasing} function of radius.  As discussed in \citet{Lightman78},
this arises because the change in energy per orbit, $\epsilon$, is
constant, while the energy of a star scales as $r^{-1}$.  At greater
clustocentric distances stars therefore require fewer orbits to change
$E^2$ by of order itself.  The increasing period with increasing
distance is not sufficient to counteract this trend.

The addition of a dark halo modifies this behavior, such that larger
dark matter contributions result in longer relaxation times.  For
sufficiently large dark matter fractions, the relaxation time will
exceed the age of the Universe.  Figure \ref{fig:trelax} includes an
upper bound provided by the age of the Universe in units of the
half-mass relaxation time of one GC we will consider in the next
section, MGC1.  For this cluster, the relaxation time in the halo does
not exceed the age of the Universe at $<6R_h$, for $M_{\rm
  DM}/M_\ast=100$ and at $R<30R_h$ for $M_{\rm DM}/M_\ast=10$.  The
very outer stellar halo of isolated GCs must be interpreted with these
facts in mind.

\citet{Lightman78} provided a simple derivation of the stellar density
profile in the GC halo.  In a steady state the net stellar flux
through a spherical shell of radius $r$ must be constant, which
implies that $n(r)r^3/t_r=\,$const.  This formula for $n(r)$, when
combined with Equation \ref{e:tr}, reproduces the stellar density
profile derived earlier in this section (Equation \ref{eqn:nr}) in the
limit where $E\sim0$.  This is not surprising because the principal
assumption in both derivations is the existence of a steady state in
the halo.

\subsection{Results}

\begin{figure}[!t]
\center
\resizebox{3.4in}{!}{\includegraphics{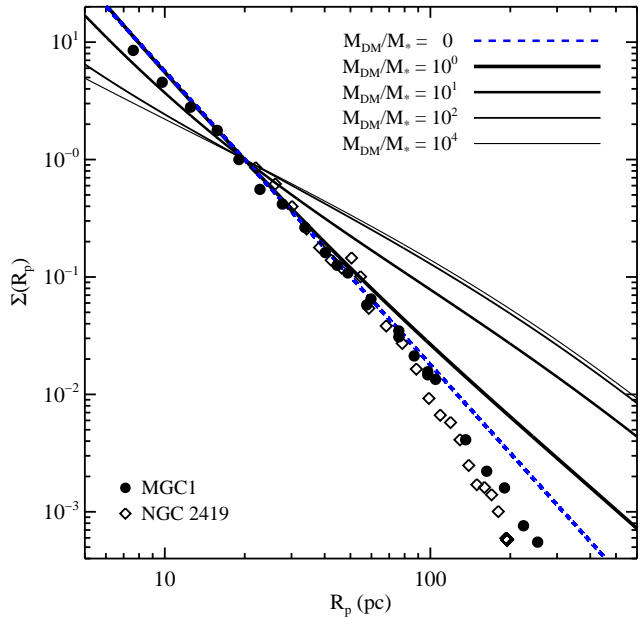}}
\caption{Stellar surface density profiles normalized to the surface
  density at 20 pc.  Our models, which include a stellar component of
  mass $M_\ast$ embedded within a dark halo of mass $\Mdm$ are shown
  as lines for a range of mass ratios.  These models are compared to
  data from the GC MGC1 located in the outer halo of M31
  \citep{Mackey10} and the GC NGC 2419 located in the outer halo of
  the MW \citep{Bellazzini07}.  Data are only plotted for $R_p>R_h$.
  The blue dashed line has a logarithmic slope of $-2.5$ and is the
  predicted surface density profile for a pure stellar system.}
\label{fig:f2}
\vspace{0.1cm}
\end{figure}

As mentioned in the Introduction, most ancient GCs are on orbits that
would likely have resulted in severe stripping of an extended dark
halo, were they originally embedded in such halos.  GCs at large
galactocentric distance, in contrast, orbit within very weak tidal
fields, and so one may expect these objects to have retained their
dark halos, if they ever had them.

Two GCs are particularly noteworthy in this regard: NGC 2419 in the MW
and MGC1 in M31.  NGC 2419 resides at 90 kpc from the center of our
Galaxy, has a half-mass and tidal radius of 20 pc and 230 pc,
respectively, and a $V-$band luminosity of $5\times10^5\,L_\Sol$
\citep{Harris96}, which implies a total stellar mass of
$\approx10^6\Msun$.  \citet{Bellazzini07} recently measured the
stellar surface density of NGC 2419 to 200 pc.  The core and half-mass
relaxation times of this GC are 9 and 35 Gyr, respectively.

\citet{Mackey10} recently measured structural and photometric
properties of MGC1, from which we have learned the following. MGC1
resides at approximately 200 kpc from M31, and is therefore the most
isolated GC known in the Local Group.  It has a $V-$band luminosity of
$4\times10^5\,L_\Sol$ and thus a stellar mass of $\approx 10^6\Msun$,
a half-mass radius of $\approx7.5$ pc, and an indeterminate tidal
radius.  Mackey et al. have measured the stellar surface density for
MGC1 out to an impressive 900 pc.  We can estimate the core and
half-mass relaxation times of MGC1 by scaling the relaxation times of
NGC 2419 by the 3/2 power of the ratio of their half-mass radii.
Doing so yields core and half-mass relaxation times of 2 and 8 Gyr,
respectively.

In Figure \ref{fig:f2} we compare the observed stellar surface density
profiles of NGC 2419 and MGC1 to our model density profile for several
values of the dark halo-to-stellar mass ratio, $\Mdm/M_\ast$.  We have
fixed $r_s=250$ pc and hence $c=4$ for simplicity.  Such a low value
of $c$ is expected for low mass halos that formed at high redshift
\citep{NFW97}.  Data are only shown for $R_p>R_h$.  On scales smaller
than roughly the half-mass radius our assumptions break down \citep[as
demonstrated by direct $N-$body simulations;][]{Baumgardt02}.

Over the range $10\lesssim R_p\lesssim100$ pc the data are consistent
with the predictions for a pure stellar system; models with a massive
extended dark halo are strongly disfavored.  On larger scales
deviation between the data and models is apparent, which may be due to
tidal stripping or the ongoing assembly of the outer stellar halo.

Our model assumes that a steady state has been achieved in the stellar
halo via two-body relaxation effects.  MGC1 has a half-mass relaxation
time shorter than the age of the Universe, and so our technique can be
readily applied to this system.  Indeed, in the absence of a dark halo
the effective relaxation time decreases with radius (Figure
\ref{fig:trelax}) and so we expect a well-developed stellar halo
around MGC1.  For models with massive dark halos (i.e., large
$\Mdm/M_\ast$), our results strictly apply only to the inner several
half-mass radii --- at larger radii the effective relaxation time
becomes longer than the age of the Universe.  However, even within
$1<R/R_h<2$ the data strongly favor models with minimal dark matter
halos.  Moreover, the data at larger radii are naturally explained by
assuming that the stellar halo is fully populated by relaxation
effects in the absence of an embedded dark halo. 

Conclusions regarding NGC 2419 must be made with greater caution than
MGC1, since NGC 2419 has a present $t_r$ that is longer than the age
of the Universe.  In this case it is less clear that our model should
apply at all.  We are left only with the striking agreement between
the density profile of the halo of this GC and the model prediction
that includes no dark halo.  This strongly suggests, but does not
rigorously demonstrate, that NGC 2419 is not surrounded by a massive
dark halo.

Our results are consistent with \citet{Baumgardt09}, who concluded
that if a dark halo surrounds NGC 2419, it cannot be more massive than
$10^7\Msun$ (this is equivalent to a limit of $\Mdm/M_\ast<10$ for
this GC).  These latter results were based on the measured velocity
dispersion profile of NGC 2419 over the range $10\lesssim R_p \lesssim
60$ pc.

\section{Discussion}
\label{s:res}

In the previous section we argued that the observed stellar surface
density profiles of the GCs NGC 2419 and MGC1 place strong constraints
on the existence of extended dark halos surrounding these GCs.  The
data are consistent with no dark halo, and a firm upper limit on the
dark halo mass-to-stellar mass ratio is $\Mdm/M_\ast<1$.  The
conclusions are strongest for MGC1 because it, unlike NGC 2419, has a
core relaxation time much less than the age of the Universe.

This upper limit effectively rules out the possibility that these GCs
formed at the center of their own dark halos, under the assumption
that these GCs have evolved in weak tidal fields throughout their
lifetimes.  This assertion is based on the following argument.  If
these GCs {\it did} form within their own dark halos and subsequently
experienced little tidal stripping, then the smallest possible value
for $\Mdm/M_\ast$ would be $(1-f_{b})/f_{b}$ where $f_{b}$ is the
universal baryon fraction.  Constraints from the cosmic microwave
background imply $f_{b}=0.17$ \citep{WMAP5}, and so $\Mdm/M_\ast>5$.
Of course, less than 100\% star formation efficiency, which is
expected, would only increase this lower limit.  Our upper limit of
$\Mdm/M_\ast<1$ therefore strongly suggests that these GCs {\it did
  not} form within their own dark halos.

Observations of the outer stellar profile of isolated GCs are very
sensitive to a dark halo because a dark halo, were it to exist, should
have a half-mass radius much larger than the GC stellar half-mass
radius.  This fact also explains why it has historically been so
difficult to obtain strong constraints on the presence of a dark halo
with kinematic data, even with data extending to several tens of
parsecs \citep[e.g.,][]{Lane10}.  An NFW dark matter halo with a
virial mass of $10^8\Msun$ has a mass of only $10^6\Msun$ within 50
pc, assuming $c=2$ (or within 10 pc assuming $c=20$).  For NGC 2419,
which has a stellar mass of $\approx10^6\Msun$, the presence of such a
halo would be very difficult to distinguish from the uncertain
corrections required to account for low mass stars and stellar
remnants, based on data that only extends to several tens of pc.

In recent years it has become clear that most, if not all GCs harbor
internal spreads in the abundance of light elements, including CNO,
Na, Mg, and Al \citep[see][for a review]{Gratton04}.  Several authors
have appealed to GC formation at the center of extended dark halos to
account for these puzzling observations \citep[e.g.,][]{Freeman93,
  Bekki06, Bekki07, Boker08, Carretta10c}.  One of the advantages of
forming GCs at the center of massive dark halos is that they are much
less susceptible to ram pressure stripping, and, the argument goes,
are therefore better able to retain the gaseous material necessary to
account for the observed internal abundance spreads.  As discussed in
\citet{Conroy11b}, this line of reasoning is likely incorrect because
the formation environments of the ancient GCs differed substantially
from their present day environment.  The results in this work provide
strong independent confirmation that indeed GCs which harbor multiple
stellar populations do not (or need not) form within extended dark
halos.

While the current evidence disfavors typical GCs from having formed at
the center of their own dark halos, there is some reason to suspect
that perhaps some of the most massive GCs did indeed form in this way.
M54 is the most striking example, as it resides at the center of the
disrupting Sagittarius galaxy, and will in the future likely orbit
freely through the Galaxy \citep[although recent evidence suggets that
M54 resides at the center of Sagittarius because of dynamical
friction, not because it formed there; see][for
details]{Bellazzini08}.  Other candidates for this formation mechanism
include $\omega$Cen, M22, NGC 1851, and G1 in M31, all of which show
internal spreads in the Fe-peak elements.  These GCs must have formed
in deep potential wells in order to retain the Fe generated from type
Ia SNe.  Nuclear star clusters may be the precursors of these massive
GCs.  The most massive GCs in external galaxies also appear to be
self-enriched in Fe \citep{Strader08, Bailin09}, although the fact
that their photometric properties join seamlessly with the less
massive clusters suggests that GCs of all masses share a common origin
unrelated to dark halos.  Detailed simulations will be necessary to
conclude whether or not GCs can self-enrich in SNe products without
surrounding dark halos.

\citet{Olszewski09} recently reported the discovery of a 500 pc
stellar halo surrounding the GC NGC 1851.  Over the projected radial
range of $50-250$ pc, these authors find a projected stellar density
profile of $\Sigma\propto r^{-1.24\pm0.66}$.  This measured profile
agrees remarkably well with models that include a massive dark halo
($\Mdm/M_\ast>10^2$), which predict a logarithmic slope of $-1.4$ over
the same radial range.  NGC 1851 currently resides only 17 kpc from
the Galactic center and, according to \citet{Olszewski09}, has a
period of 0.4 Gyr and a perigalacticon of only 5 kpc.  The
interpretation of the density profile of weakly-bound stars in this
cluster is therefore greatly complicated by the stronger tidal fields
it experiences and the effect of disk shocking as it crosses the MW
disk five times per Gyr.  The lack of any tidal tails is also peculiar
given its orbit.  As noted above, NGC 1851 shows evidence for an
internal spread in Fe abundance \citep{Carretta10d}, and so is a
potential candidate for being the remnant of a disrupted dwarf galaxy.
Future work on the orbit and stellar population of this cluster may
reveal important clues regarding its formation.  Radial velocity
measurements would be especially valuable, as they should be able to
distinguish between a stellar halo formed from tidal effects and one
formed from loosely bound stars on radial orbits.

Recently, \citet{Cohen10} measured iron and calcium abundances of
stars in NGC 2419.  These authors report the discovery of an internal
spread in Ca abundances in this cluster, but no spread in Fe.  If
confirmed, this result suggests that NGC 2419 was able to retain type
II SNe ejecta, which is difficult to understand unless this cluster
was once embedded within a much deeper potential well than it is
currently.  It could of course be the case that the stars in NGC 2419
simply formed from a chemically heterogenous molecular cloud, or that
the cluster contained many more stars at birth.  As with NGC 1851,
future work on the abundance variations of the stars within NGC 2419
and a detailed analysis of its orbit will provide essential clues into
the origin of this puzzling GC.

We conclude by recalling a central assumption in the present work:
that outer halo GCs have evolved in isolation throughout their
lifetimes.  Unless these GCs formed in intergalactic space, they
likely once resided within larger protogalactic fragments that have
since been tidally destroyed.  We can say little with confidence
regarding the influence of the birth environment on the structure of
these GCs.  Mass lost from these young GCs during their first $\sim1$
Gyr of evolution would result in an expansion of the system due to the
loss of binding energy \citep{Kroupa02, Marks10}.  These effects
complicate the interpretation of the outer stellar halo of the GCs NGC
2419 and MGC1.  Nonetheless, the tenuous nature of their stellar halos
and the observed similarity in their radial profiles strongly
suggests that they are being continuously populated by two-body
relaxation effects.


\acknowledgments 

We thank Dougal Mackey for providing his data on MGC1, Jay Strader for
fruitful conversations, and Dougal Mackey and Scott Tremaine for
comments on an earlier draft.  The referee is thanked for insightful
comments that improved the quality of the manuscript.  This work made
extensive use of the NASA Astrophysics Data System and of the {\tt
  astro-ph} preprint archive at {\tt arXiv.org}, and was supported in
part by NSF grants AST-0907890 and AST-0707731 and NASA grants
NNX08AK43G and NNA09DB30A.


\end{document}